\documentclass[12pt]{article}
\usepackage[brazilian]{babel} 
\usepackage[T1]{fontenc} 
\usepackage{ae} 
\usepackage[latin1]{inputenc}
\usepackage{graphicx}
\usepackage{amsfonts}
\usepackage{amssymb}
\usepackage{epsfig}
\usepackage{amsmath}

\begin{document}

\title{A simple mathematical model underlying Keller's individualized teaching method} 

\centerline{\bf A simple mathematical model underlying Keller's individualized teaching method}
\vspace{0.2truein}

\centerline{\small Um modelo matem\'{a}tico simples subjacente ao método de ensino individualizado de Keller} 
\vspace{0.2truein}

\centerline{\footnotesize  Danilo T. Alves$^{\mbox{(a,b,c)}}$\footnote{danilo@ufpa.br}, 
Nelson P. C. de Souza$^{\mbox{(d)}}$ \footnote{npcoelho@yahoo.com},
\footnotesize{Silvio C. F. Pereira Filho$^{\mbox{(e)}}$\footnote{silviocfilho@ufpa.br},}} 
\vspace{0.015truein}
\centerline{\footnotesize (a) {\it Faculdade de F\'{i}sica, Universidade Federal do Par\'{a}, Bel\'{e}m, PA, Brazil}}
\centerline{\footnotesize (b) {\it 
Instituto de Educa\c{c}\~{a}o Matem\'{a}tica e Cient\'{i}fica, Universidade Federal do 
Par\'{a}, Bel\'{e}m, PA, Brazil}}
\centerline{\footnotesize (c) {\it 
	Centro de F\'{i}sica, Universidade do Minho, P-4710-057, Braga, Portugal}}
\centerline{\footnotesize (d) {\it 
Escola de Aplica\c{c}\~{a}o, Universidade Federal do Par\'{a}, Bel\'{e}m, Brazil}}
\centerline{\footnotesize (e) {\it 
Faculdade de Ci\^{e}ncias Naturais, Campus Universit\'{a}rio do Maraj\'{o}-Breves,}} 
\centerline{\footnotesize\it{Universidade Federal do Par\'{a}, Breves, Brazil}}

\begin{abstract}
	Keller's article entitled ``Good-bye teacher...'' [J. Appl. Behav. Anal. \textbf{1}, 79-89 (1968)]
	was fundamental for the development and dissemination of Keller's Personalized System of Instruction 
	(PSI), which was one of the central issues in the discussions on psychology and education
	in the 1970s and 1980s, and nowadays has attracted attention in the context of the 
	increasing use of online education.
	Belonging to a class of approaches usually named as mastery learning, PSI and
	modified PSI courses (Keller-type courses) present several interesting results, such as final grade distributions
	where the majority of students achieve the highest grades. 
	Here, we present a simple mathematical model underlying Keller-type individualized teaching methods,
	describing, in terms of average characteristic parameters, the time evolution of 
	the distribution of students per unit of content, and that most students achieve the highest grades at the end of the course. 
	By applying this model to a real case of an  introductory electromagnetism Keller-type course, 
	we obtained its characteristic parameters with which we showed good agreement between the predictions and observations.
	The model presented here results in a simple formula, which is very accessible for use by a wide audience interested in planning or investigating Keller-type or other mastering learning methods.
\end{abstract}
\maketitle

\section{Introduction} 
\label{introduction}
In his paper of 1968 \cite{KELLER_68}, Keller describes his instructional plan (also named Personalized System of Instruction - PSI) 
as one through which the student can move at his own pace, showing mastery, not being forced to go ahead until he is ready. 
Like Bloom's learning for mastery \cite{Bloom-Eval-Comment-1968,Bloom-Educ-Rese-1984,Guskey-1985}, the PSI
belongs to a class of approaches usually named as mastery learning.
In Keller's scheme ``...the student go ahead to new material only after demonstrating mastery of that which preceded, before to go to the next one'' \cite{KELLER_68}.
The mastery learning emerged as an alternative to traditional courses - 
those where students spend most of their time in classroom attending lectures,
and assessments are only meant to score the success obtained at the course - and it has shown
``positive effects on the examination performance of students in colleges, high schools,
and the upper grades in elementary schools'' \cite{Kulik-1990}.
In a Keller course, assessments are not only a measure of success, 
but are also formative. An assessment is considered to be formative if it ``provides students opportunities to revise and hence improve the quality 
of their thinking and learning'' \cite{BRANSFORD_99}. 
The formative aspect of the assessments in PSI is enhanced by the systematic feedback 
received by the students right after these assessments \cite{KELLER_68,SHERMAN_92}.
Both formative assessments and systematic feedback are presently considered fundamental to learning \cite{BRANSFORD_99}.

The PSI originally started as a result of the application of the reinforcement theory, 
according to which the instruction process must be based on presentation, performance and consequences, 
maximizing the frequency of reinforcement and reducing the aversive consequences of errors \cite{SKINNER_1,SKINNER_2}. This led to the main PSI features, 
among them: mastery, self-pacing, linear small-step sequenced materials (each one related to a unit of content), repeated testing and immediate feedback \cite{SHERMAN_92}.
In a Keller course, students spend most of their time in classroom doing assessments, receiving feedback 
or studying the instructional material: ``instead of responding passively  to  a  lecture,  students  must  actively  read,  study,  
and  respond  in writing  to  questions  over  textual  materials'' \cite{BUSKIST-1991}. 
In PSI, lectures are ``...vehicles of motivation, 
rather than sources of critical information'' \cite{KELLER_68} and are typically short \cite{BUSKIST-1991}.
An interesting result of the Keller method is the ``production of a grade distribution that is upside down'' \cite{KELLER_68}, which means that, instead of a typical bell-shaped curve for the final grades found in traditional courses \cite{Guskey-1985}, 
in PSI courses a higher percentage of students reaches the highest final grade.

Keller's method has been applied and discussed in the context of several disciplines \cite{KULIK_74},
including Physics \cite{Moreira-RBF-1973,Bezerra-RBF-1973,Bezerra-RBF-1974,Dionisio-RBF-1975,alves-filho-souza-elias-RBEF-2011}, and recent advances in the technology of information are being considered for implementing and improving the Keller course \cite{EYRE_07,FOX_04}. 
Although evidences derived from observations indicate success of the Keller scheme in many aspects,
it began to decline in popularity during the 1980's.  
Among the causes,are
criticisms on the behaviorist approach of the method, as well as
the high workload associated with the implementation of a Keller course \cite{EYRE_07,Silberman-1978}.

In the present paper, we present a simple mathematical model to describe 
a PSI \cite{KELLER_68} or a modified PSI course (Keller-type course) \cite{alves-filho-souza-elias-RBEF-2011}.
With this model, we can estimate the evolution of the distribution of students per unit of content,
the presence of the upside down effect in the grades and under which
situations this effect occurs. 
We compare our theoretical predictions with results from a real application of a Keller-type method found in the literature \cite{alves-filho-souza-elias-RBEF-2011}.

The paper is organized as follows. In Sec. \ref{exact}, we discuss a mathematical model 
to describe a very general Keller-type course. 
In Sec. \ref{app}, we propose approximations, obtaining a simple formula,
which allows the visualization of several interesting
aspects of Keller-type courses.
In Sec. \ref{app-real}, we apply this model to a real case of an introductory level electromagnetism Keller-type course, described in Ref. \cite{alves-filho-souza-elias-RBEF-2011}.
In Sec. \ref{aplicacao-pratica}, we exemplify how our model can be used by a teacher 
interested in planning and improving a Keller-type course.
In Sec. \ref{final}, we make our final considerations.

\section{A general model}
\label{exact}

In traditional courses, concepts and materials are divided into $N_u$ units of content, which
``...correspond, in many cases, to chapters in the textbook used in teaching'' \cite{Guskey-1985}.
Usually, $N_a$ (being $N_a\geq N_u$) tests are administered to the students. Usually each test covers the content of a given unit,
so we can say that in traditional courses $N_a \approx N_u$. To the
teacher, a test is ``...an evaluation device that determines who learned those concepts well and who did not'' \cite{Guskey-1985}.
To the students, each test means ``...the end of instruction on the unit and the end of the time they need
to spend working on those concepts'' and most of the time ``...the only chance to demonstrate what they learned'' \cite{Guskey-1985}.
``After the test is administered and scored, marks are recorded in a grade book, and instruction begins on the next
unit, where the process is repeated'' \cite{Guskey-1985}.
The structure of a course according to the Keller plan \cite{KELLER_68} is very different and can be modeled as follows. 

The course content is divided into $N_u$ units, organized in a definite
numerical order ($1,...,N_u$), and the students have to
show mastery in each unit by passing assessments (tests, works, etc.). 
Let us consider that there are $N_a$ total opportunities of such assessments related to these $N_u$ units,
so that, if the student fails to pass an assessment on the first opportunity, there can be other opportunities to be used.
It can be noticed that in typical Keller courses $N_a \gg N_u$, whereas in traditional courses $N_a \approx N_u$.
In addition, in a Keller course it can be applied $N_e$ final examinations (but commonly $N_e=1$), each one of them applied at the same time
for all students.
Usually, a certain percentage $\lambda_u$ of the course grade is based on the number of units successfully completed during
the term, a percentage $\lambda_f$ is based on the final examinations, 
whereas the percentage $1-(\lambda_u+\lambda_f)$ is associated with exercises (laboratory works, etc.).
This means that  
there can have a minimal number $N_c$ of units of content to be
successfully completed by the students to get approval. 
Here, we call Keller-type to any course
whose content is divided into $N_u$ units, organized in a definite
numerical order ($1,...,N_u$), with the students having to
show mastery in each unit before going to the next one, having a total
of $N_a$ opportunities of assessments,
from which a number $N_c$ of units of content must be successfully completed to get approval. 
The mathematical model presented here underlies Keller-type courses,
which include the original Keller plan \cite{KELLER_68}.

\begin{figure}
\centering                  
\includegraphics[scale=0.55]{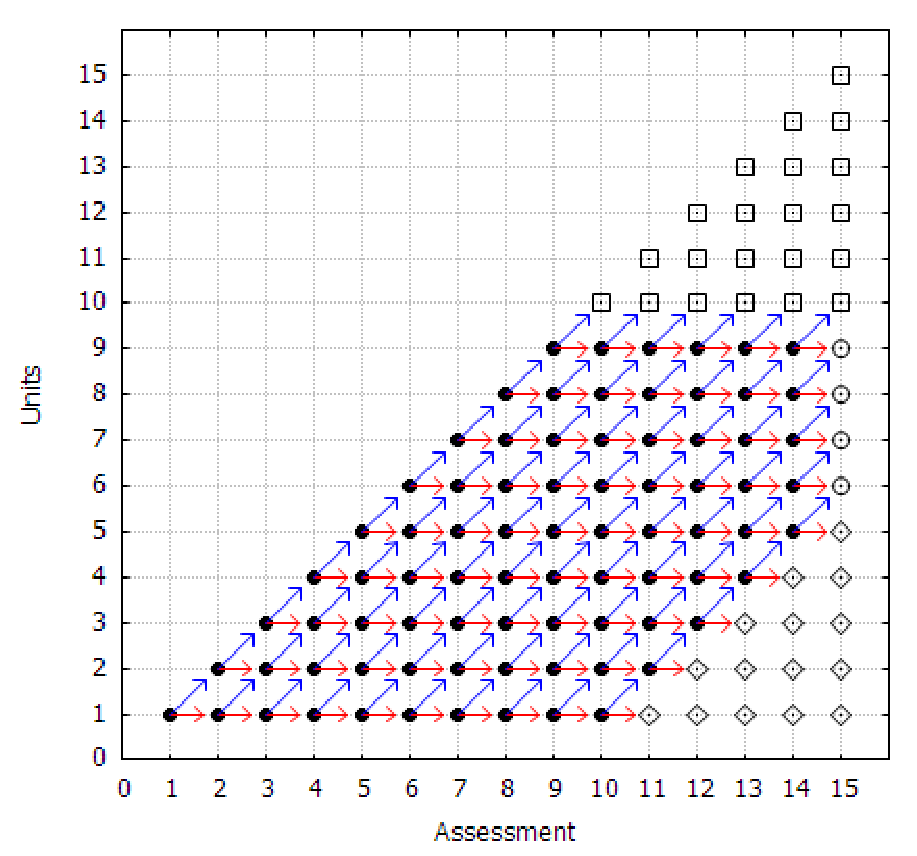}                                                                                  
\caption{The scheme of assessments of a Keller course with $N_u=9$, $N_c=5$ and $N_a=14$. In the horizontal axis we set a number for each  opportunity of assessment [indicated by the first index, $i$, in $N_{(i,j)}$ along the text], whereas 
in the vertical axis  we set a number for each unit of content
[indicated by the second index, $j$, in $N_{(i,j)}$].
The points indicated by dark circles mean possible real situations 
of assessment. The points indicated by squares, diamonds and
circles mean virtual situations.
}
\label{keller-plan}
\end{figure}

Let us consider a Keller-type course,
with $N_s$ students.
In Fig. \ref{keller-plan} we represent in a Cartesian plane the scheme of assessments of a Keller course,
with $N_u=9$, $N_c=5$ and $N_a=14$. 
In the horizontal axis we set a number for each opportunity of assessment, whereas 
in the vertical axis  we set a number for each unit of content. In this manner, the point
$(1,1)$ means that in the first opportunity of assessment the students can be trying to pass the
assessment related to the first unit of content. The point
$(2,1)$ means that in the second opportunity of assessment, students are trying to pass the
assessment related to the first unit of content, whereas the point $(2,2)$ shows that in the
same opportunity of assessment there are students trying to pass the
assessment related to the second unit of content. 
The other points have analogous meaning.
Considering $N_{(i,j)}$ as the number of students in the  $i$th opportunity of assessment, trying to pass 
the assessment related to the $j$th  unit of content, one can directly establish relations to describe the propagation
of the number of students from a point to another, so that $N_{(i,j)}$ can be written as
\begin{equation}
	N_{(i,j)}=\begin{cases}
		N_{s}\;(i=j=1),\\\
		\alpha_{(i-1,1)}N_{(i-1,1)}\;(i\geq2;j=1),\\\
		\beta_{(i-1,i-1)}N_{(i-1,i-1)}\;(i=j>1),\\\
		\alpha_{(i-1,j)}N_{(i-1,j)}+\beta_{(i-1,j-1)}N_{(i-1,j-1)}\\
		(i\geq j+1>2),
	\end{cases}
	\label{N}
\end{equation}
where $i$ and $j=1,2...$, the $\alpha$ coefficients are related to 
the failure in passing an assessment, whereas the $\beta$ coefficients are related to 
the success (mastery in a given unit of content). 
In Fig. \ref{keller-plan}, the horizontal (red) arrows represent the propagation
- mediated by the $\alpha$ coefficients - of the number of students that failed a given assessment,
whereas the diagonal (blue) arrows represent the propagation
- mediated by the $\beta$ coefficients - of the number of students that obtain success.
The points indicated by dark circles mean possible real situations 
of assessments, whereas the points indicated by squares, diamonds and circles
are virtual points, which means that there are no real assessment related to these points.
These virtual points are inserted because they will be useful (see Sec. \ref{app}) in 
the analysis of the time evolution of the number 
of students in each unit: those who have already concluded all real $N_u$ units will ``occupy'' virtual
units (squares); those students that have not concluded
the minimal number $N_c$ of units will occupy the diamond points;
and those who concluded between $N_c$ and $N_u$ units occupy the circle points.  
This can be illustrated through the following examples: 
(i) students at the point $(14,9)$ who pass this assessment complete all units, having
no more assessments to pass, reach the ``virtual'' point $(15,10)$;  
(ii) those at the point $(14,9)$ and that do not pass this assessment go to the ``virtual'' point
$(15,9)$;
(iii) and those at the point $(14,5)$ and that pass this assessment reach the ``virtual'' point
$(15,6)$.
\begin{figure}[htb]
	\centering                  
	\includegraphics[scale=1.00]{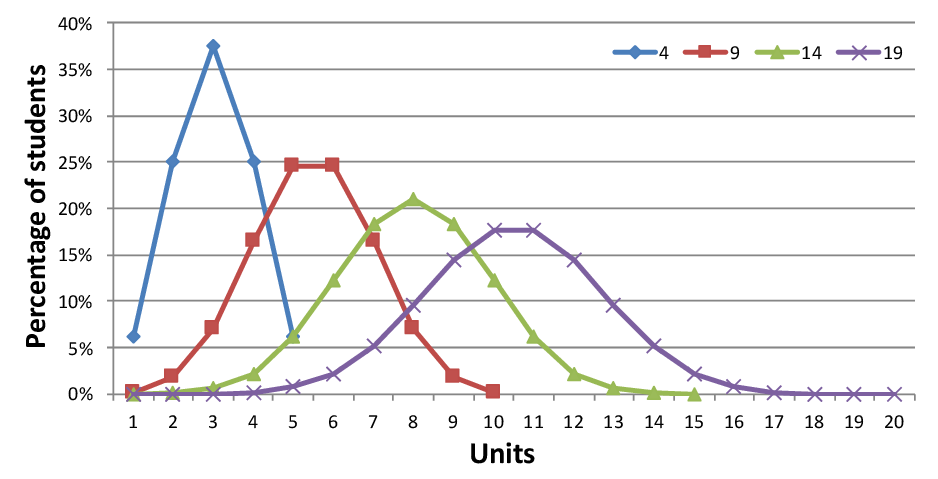}                                           
	\caption{
		For the case $(\alpha,\beta)=(1/2,1/2)$, the horizontal axis gives the number of each unit of content,
		whereas the vertical axis exhibits the percentage of students 
		($N_{(i,j)}/N_s$) who are able to do the assessment correspondent to each unit.
		After 4 assessments, the situation is described by the blue line (diamonds),
		which shows $N_{(5,j)}/N_s$
		for $j=1..5$; after the 9th assessment, the red line
		(squares) shows $N_{(10,j)}/N_s$
		for $j=1..10$; after the 14th assessment, $N_{(15,j)}/N_s$ for $j=1..15$
		is described by the green line (triangles); and after the 19th assessment,
		the situation is described by $N_{(20,j)}/N_s$ for $j=1..20$, which is indicated 
		by the purple line (crosses).
	}
	\label{evolucao-meio-meio}
\end{figure}
\begin{figure}[!hbt]
	\centering                  
	\includegraphics[scale=1.3]{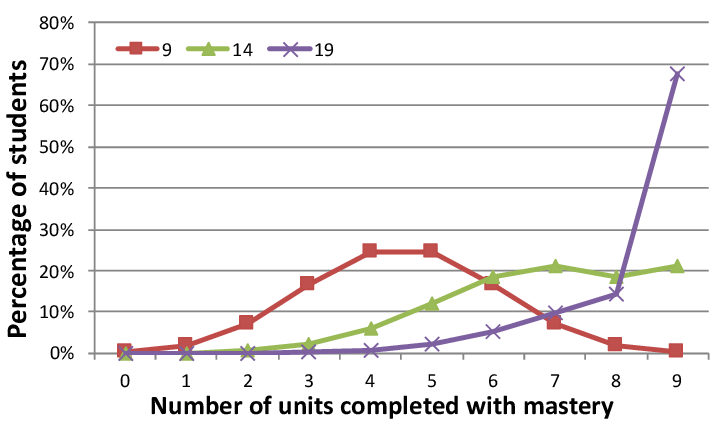}                                           
	\caption{The horizontal axis shows the number of units completed with mastery. 
		The vertical axis exhibits the percentage of students.
		The cases considered are defined by $(\alpha,\beta)=(1/2,1/2)$, $N_u=9$ and for the following 
		values of $N_a$: for $N_a=9$ (squares on the red line); for $N_a=14$ (triangles on the green line);
		for $N_a=19$ (crosses on the purple line).
	}
	\label{inversao}
\end{figure}
\begin{figure}[!hbt]
	\centering                  
	\includegraphics[scale=1.3]{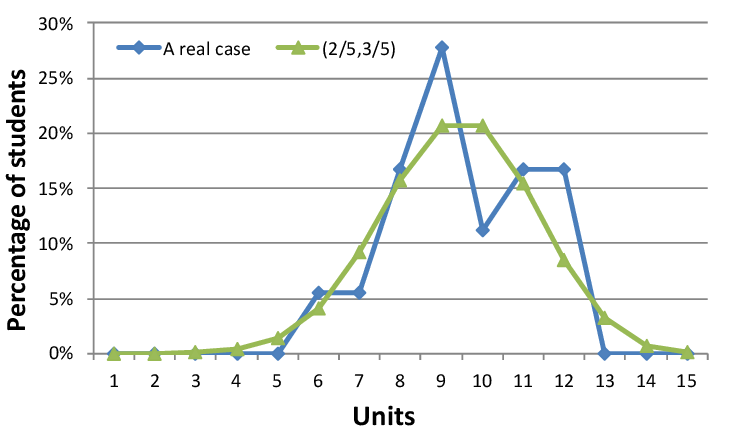}                                           
	\caption{The horizontal axis shows the number of each unit of content,
		whereas the vertical axis exhibits the percentage of students 
		($N_{(i,j)}/N_s$) who are able to do the assessment correspondent to each unit,
		after the 14th assessment, for the case $(\alpha,\beta)=(2/5,3/5)$ (triangles on the green line), and a real case described in Ref. \cite{alves-filho-souza-elias-RBEF-2011} (diamonds on the blue line).}
	\label{comparacao-real-teoria}
\end{figure}
\begin{figure}[!hbt]
	\centering                  
	\includegraphics[scale=1.3]{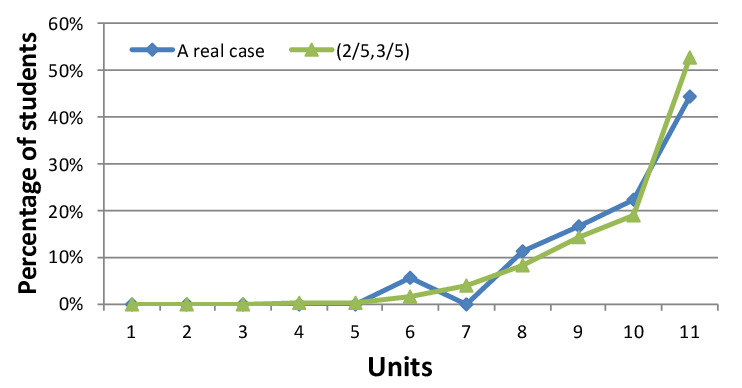}                                           
	\caption{The horizontal axis shows the number of units completed with mastery. 
		The vertical axis exhibits the percentage of students.
		The cases considered are $(\alpha,\beta)=(2/5,3/5)$ (triangles on the green line), and a real case described in Ref. \cite{alves-filho-souza-elias-RBEF-2011} (diamonds on the blue line).
		We considered $N_u=11$ and $N_a=16$.}
	\label{inversion-real-theory}
\end{figure}

\section{An approximate model}
\label{app}

Equation (\ref{N}) is very general, and a direct description underlying
any Keller-type course.
If all coefficients $\alpha_{(i,j)}$ and $\beta_{(i,j)}$ are known,
equation (\ref{N}) predicts the number of students $N_{(i,j)}$ in the $i$th opportunity of assessment, trying to pass the assessment related to the $j$th unit of content, in each moment of the course.
The various elements that can affect such a course,
for example, the number and quality of proctors and assistants,
number of sessions, the quality of textbooks and other study materials, among others, 
can influence on a best (decreasing $\alpha$ and increasing $\beta$ parameters) or worse (increasing $\alpha$ and decreasing $\beta$) student performance.
Specifically, if students make more use of assessment opportunities, $\alpha$ decreases and $\beta$ increases;
on the contrary, taking less advantage of assessment opportunities, $\alpha$  increases and $\beta$ decreases.
In other words, all these characteristics can be encapsulated into the parameters $\alpha_{(i,j)}$ and $\beta_{(i,j)}$.

We point out that, in real situations, the use of equation (\ref{N}) may involve a large number of
parameters $\alpha_{(i,j)}$ and $\beta_{(i,j)}$, requiring some computational effort.
Therefore, certain approximations with the intention of generating a more simplified model need to be considered.
For example, a simplification is obtained by assuming that, for
a given $j$th unit, the probability of success in passing does not depend on which is the $i$th opportunity of assessment used by the student \cite{Bezerra-RBF-1974}.
In other words, $\beta_{(i,j)}=b_j$, where $b_j$ is a value that only depends on $j$.

In order to obtain a simpler formula than those found in Ref. \cite{Bezerra-RBF-1974}, 
but which already captures the essential features underlying Keller-type individualized teaching methods,
and also can be used by a wide audience interested in 
planning or investigating Keller-type courses, here we introduce additional approximations.
From now on we focus on describing our proposal for an approximate average behavior of $N_{(i,j)}$ given in equation (\ref{N}).
We consider that:
(i) there is no dropping out of the course, so that the number $N_s$ of students at any time will be always the same;
(ii) in each opportunity, a given student try to pass in just one of the $N_u$ units of content.
Considering these approximations, we can write
\begin{eqnarray}
	{\alpha}_{(i,j)}+{\beta}_{(i,j)}&=& 1.
	\label{milho_0}
\end{eqnarray}
Hereafter, we will consider the approximate model where, for all $i$ and $j$,  
\begin{equation}
	\alpha_{(i,j)}=\alpha,\; \beta_{(i,j)}=\beta, 
	\label{m7}
\end{equation}
with $\alpha$ and $\beta$ constants. 
This can be interpreted in two ways: the probability (or difficulty) of passing an assessment is exactly the same
for all assessments (which can be considered unrealistic);
$\alpha$ and $\beta$ are average (taken over all units) 
characteristic parameters of a Keller-type course, and this is the interpretation we will adopt 
throughout the text.

Using equation (\ref{m7}) in equation (\ref{N}), 
after some manipulations (see Appendix \ref{ap:deducao}), we have 
\begin{equation}
	N_{(i,j)}={\frac { \left( i-1 \right) !\,{\alpha}^{i-j}{\beta}^{j-1}}{ 
			\left( j-1 \right) !\, \left( i-j \right) !}}N_{s},
	\label{N-teorico}
\end{equation}
where the factor multiplying $N_{s}$ is the Bernoulli probability function,
and, as already mentioned, $N_{(i,j)}$  is the number of students 
at the position $(i,j)$ of the diagram in Fig. \ref{keller-plan}.

To illustrate an use of equation \ref{N-teorico},
let us consider, for instance, the case $(\alpha,\beta)=(1/2,1/2)$.
For this situation, we get the time (given in terms of the number of assessments already done) evolution of the class as shown in Fig. \ref{evolucao-meio-meio}, which exhibits the situations of the class
after the 4th, 9th, 14th and 19th assessments. 
For instance, after $4$ assessments (blue line with points indicated by diamonds), the situation
is described by $N_{(5,j)}/N_s$, and the figure shows that 
$N_{(5,1)}/N_s\approx 6\%$ of the students need to redo the assessment of the 1st unit,
whereas $N_{(5,2)}/N_s\approx25\%$ are able to do the assessment of the 2nd unit,
$N_{(5,3)}/N_s\approx38\%$ are able to do the assessment of the 3rd unit,
$N_{(5,4)}/N_s\approx25\%$  of the 4th unit, and 
$N_{(5,5)}/N_s\approx6\%$ are able to do the assessment of the 5th unit.
The red line (with points indicated by squares)
shows that  after the 9th assessment and for $N_u=9$, among other results,
$N_{(10,9)}/N_s\approx 1.8\%$ of the students are able to do the assessment of the last 9th unit and $N_{(10,10)}/N_s\approx 0.2\%$ are 
at the ``virtual'' unit 10, which means that these students have already shown mastery
in all 9 units. 
The purple line (with points indicated by crosses)
shows, after the 19th assessment and for $N_u=9$, among other results,
that the sum over the virtual units $\sum_{j=10}^{20}N_{(20,j)}/N_s\approx67.6\%$ gives
the part of the students that have already shown mastery in all the $N_u=9$ units of content.
The lines blue (diamonds), red (squares), green (triangles) and purple (crosses) exhibit
the behavior of a ``wave packet'' propagating from the left to right in Fig. \ref{evolucao-meio-meio}.
For other values of $\alpha$ and $\beta$, we get the same behavior, 
but as $\beta$ increases (it becomes easier to pass assessments)
the peak of the wave packet propagates faster, showing that students obtain a faster progress 
(mastery) in the units of content. 

To investigate the production in Keller courses of
grade distributions that is upside down \cite{KELLER_68}, let us consider
the hypothesis that as the number of units of content in which a student 
shows mastery increases, the probability of achieving the best final grades also increases.
Then, we assume that the behavior of final grades is intrinsically related to the mastery
level reached by the students. 
Let us consider, for example, 
the model described by equation (\ref{N-teorico}) with $(\alpha,\beta)=(1/2,1/2)$, and 
also $N_u=9$ for the following situations: when $N_a=9$, $N_a=14$ and $N_a=19$.
In Fig. \ref{inversao} we observe the distribution of students versus the maximum number of the units of content 
for which they obtained mastery at the end of the course. For instance, 
the squares on the red line show that for $N_a=9$ we have
a distribution of mastery which resembles a bell-shaped curve, which is typical of traditional courses \cite{Guskey-1985}.
We interpret this proximity between the results for mastery in a Keller-type course and traditional courses as a consequence of our choice of a theoretical 
model of a Keller scheme with $N_a = N_u$, being the relation $N_a \approx N_u$ typical in traditional courses \cite{Guskey-1985}.
However, in real Keller-type courses $N_a > N_u$
and one can observe in Fig. \ref{inversao} that the distribution of mastery suffers an inversion
as $N_a$ becomes greater than $N_u$.
For $N_a=14$ (triangles on the green line), we already observe a deformation
of the ``bell-shaped'' curve.
For $N_a=19$ (crosses on the purple line), we finally observe an inversion:
a transition from a bell-shaped to an ``exponential-shaped'' curve, indicating that the majority of students obtained mastery in all $N_u=9$ units. 
Then, assuming that the behavior of final grades is related to the mastery
level reached by the students, an inversion of the mastery curve can be directly mapped into
an inversion of final grades, which corresponds to the 
upside down effect mentioned by Keller \cite{KELLER_68}.
But, we remark that the present model reveals that this inversion
does not occur for an arbitrary value of $N_a$, but,
as the number $N_a$ increases in comparison
to $N_u$, the possibility of appearance of the mentioned inversion effect enhances.

\section{Application to a real Keller-type course}
\label{app-real}

Now, we apply equation \ref{N-teorico} to a real case of an introductory 
electromagnetism Keller-type course, described in Ref. \cite{alves-filho-souza-elias-RBEF-2011}.
In Fig. \ref{comparacao-real-teoria}, we exhibit the wave packet behavior (diamonds on the blue line) found
in the real situation (discarding waivers) described in Ref. \cite{alves-filho-souza-elias-RBEF-2011}.
We also exhibit the prediction from our model (triangles on the green line), considering
the average characteristic parameters $(\alpha,\beta)=(2/5,3/5)$.
One can see that the real behavior is in good agreement with the model,
for these characteristic parameters. 
This indicates that $(\alpha,\beta)=(2/5,3/5)$ 
are the average characteristic parameters for the Keller-type course described in Ref. \cite{alves-filho-souza-elias-RBEF-2011}. 

Let us use the average characteristic parameters $(\alpha,\beta)=(2/5,3/5)$ to investigate
the mastery.
In Fig. \ref{inversion-real-theory}, we exhibit the final mastery situation, considering  $(\alpha,\beta)=(2/5,3/5)$,
and compare the theoretical prediction using these parameters (triangles on the green line) with the real situation (diamonds on the blue line) described in Ref. \cite{alves-filho-souza-elias-RBEF-2011}, 
where it was considered, effectively, $N_u=11$ and $N_a=16$. 
We highlight that the inversion of grades is in good agreement with the theoretical prediction.
This reinforces that $(\alpha,\beta)=(2/5,3/5)$ are the average characteristic parameters of
the Keller-type course considered in Ref. \cite{alves-filho-souza-elias-RBEF-2011}.

\section{Use in planning and improving a Keller-type course}
\label{aplicacao-pratica}

Let us exemplify how equation (\ref{N-teorico}) can be used, for example, by a teacher 
interested in planning and improving a Keller-type course.
Among several characteristics that may differ from one Keller-type course to another, in common is the need to
divide the content into $N_u$ units,
and define the number $N_a$ of total assessment opportunities.
If, for example, the teacher divides the content into $N_u=9$ units, and,
from previous experience, estimates that an average (over the units)
of $50\%$ of students will have success in passing each	assessment,
the parameters characterizing such a course are $(\alpha,\beta)=(1/2,1/2)$.
A question for the teacher is: how many assessment opportunities are needed for most 
students to reach mastery at the end of the course? 

Using equation (\ref{N-teorico}) (as explained in Section \ref{app}), 
for $N_a=N_u=9$, the mastery curve will be a
bell-shaped curve found in traditional courses (see the squares on the red line in Fig. \ref{inversao}) \cite{Guskey-1985}.
On the other hand, if the choice is $N_a=N_u=19$, 
the teacher can expect a transition from a bell-shaped to an ``exponential-shaped'' curve, indicating that the majority of students obtain mastery in all $N_u=9$ units (see the crosses on the purple line in Fig. \ref{inversao}), which is a typical result of a Keller-type course \cite{ KELLER_68,alves-filho-souza-elias-RBEF-2011}. 

Let us now consider that, after the application of a Keller-type course,
the data collected by the teacher show that the class did not exhibit the evolution predicted in Fig. \ref{evolucao-meio-meio}, or the mastery curve predicted in Fig.\ref{inversao}, both considering $(\alpha,\beta)=(1/2,1/2)$.
Then, from this, the teacher could find the parameters $(\alpha,\beta)$ that are in agreement
with the results obtained in the applied Keller-type course, which allows an improvement in the 
choice of the number $N_a$ of total assessment opportunities, which would be useful in other applications.
Thus, equation (\ref{N-teorico}) can be used for planning and improving a Keller-type course

\section{Final remarks}
\label{final}

Our main result is the simple formula in equation (\ref{N-teorico}),
which predicts the average behavior of the number of students  $N_{(i,j)}$ in the  $i$th opportunity of assessment, trying to pass the assessment related to the $j$th unit of content, in each moment of a Keller-type course.
$N_{(i,j)}$ is described in terms of only two parameters, the coefficient $\alpha$, which is the average probability of failing in passing an assessment, and $\beta$, which is the average probability
of success (or to get mastery in a given unit of content). 
The various characteristics of a Keller-type course, for example, the number and quality of assistants,
number of sessions, the quality of textbooks and other study materials, among others, 
can influence on a best (decreasing $\alpha$ and increasing $\beta$ parameters) or worse (increasing $\alpha$ and decreasing $\beta$) student performance, with all these characteristics encapsulated in the $\alpha$ and $\beta$
parameters.
As discussed throughout the text, this makes equation (\ref{N-teorico}) very accessible for use by a wide audience interested in planning or investigating Keller-type courses or other mastering learning courses.

An interesting result coming from our analysis based on equation (\ref{N-teorico}) is that
when the number $N_a$ of assessment opportunities is approximately the same as the number $N_u$ of units of content,
the mastery curve will be a bell-shaped curve found in traditional courses 
(red line in Fig. \ref{inversao}), whereas
as $N_a$ becomes greater than $N_u$, 
there is a change from a bell-shaped to an exponential-shaped curve
(purple line in Fig. \ref{inversao}), indicating that the majority of students obtain mastery, which is a typical result of a Keller-type course. 

\section{Acknowledgments}
\label{ack}

We thank Liane R. Tarouco, Adilson O. do Esp\'{i}rito Santo,
Olavo F. Galv\~{a}o, Benedito de Jesus P. Ferreira, C\'{e}zar A. Galera,
Tadeu O. Gon\c{c}alves, Ana Leda de F. Brino, \'{I}caro Gonzaga, Romariz S. Barros and Thiago D. da Costa, 
for valuable discussions. 
This work was partially supported by Coordena\c{c}\~{a}o de 
Aperfei\c{c}oamento de Pessoal de N\'{i}vel Superior (CAPES/Brazil).

\appendix

\section{Obtaining equation (\ref{N-teorico})}
\label{ap:deducao}

This section builds on using equation 1 in equation 2.
%
For $i=2$ and $j=1,2$, we get
\begin{eqnarray}
	N_{(2,1)}&=&\alpha N_{s}\\
	N_{(2,2)}&=&\beta N_{s}.
\end{eqnarray}
In parallel, note that we can write
\begin{equation}
	(\alpha+\beta)N_{s}=\alpha N_{s}+\beta N_{s},
\end{equation}
which, for convenience, can be written as
\begin{equation}
	(\alpha+\beta)^{2-1}N_{s}=\underbrace{\alpha^{2-1}\beta^{1-1}N_{s}}_{N_{(2,1)}}+\underbrace{\alpha^{2-2}\beta^{2-1}N_{s}}_{N_{(2,2)}}.
	\label{con-1}
\end{equation}
From equation (\ref{con-1}), we identify $N_{(2,1)}$ and $N_{(2,2)}$.

For $i=3$ and $j=1,2,3$, we have
\begin{eqnarray}
	N_{(3,1)}&=&\alpha^2 N_{s},\\
	N_{(3,2)}&=&2\alpha\beta N_{s},\\
	N_{(3,3)}&=&\beta^2 N_{s}.
\end{eqnarray}
In parallel, we can also write
\begin{equation}
	(\alpha+\beta)^{2}N_{s}=\alpha^{2}N_{s}+2\alpha\beta N_{s}+\beta^{2}N_{s},
\end{equation}
which can be conveniently written as
\begin{eqnarray}
	(\alpha+\beta)^{3-1}N_{s}&=&\underbrace{\alpha^{3-1}\beta^{1-1}N_{s}}_{N_{(3,1)}}+\underbrace{2\alpha^{3-2}\beta^{2-1}N_{s}}_{N_{(3,2)}}\nonumber\\
	&+&\underbrace{\alpha^{3-3}\beta^{3-1}N_{s}}_{N_{(3,3)}}.
	\label{con-2}
\end{eqnarray}
From equation (\ref{con-2}), we identify $N_{(3,1)}$, $N_{(3,2)}$, and $N_{(3,3)}$.

For $i=4$ and $j=1,2,3,4$, we have
\begin{eqnarray}
	N_{(4,1)}&=&\alpha^3 N_{s},\\
	N_{(4,2)}&=&3\alpha^2\beta N_{s},\\
	N_{(4,3)}&=&3\alpha\beta^2 N_{s},\\
	N_{(4,4)}&=&\beta^3 N_{s}.
\end{eqnarray}
We can write
\begin{equation}
	(\alpha+\beta)^{3}N_{s}=\alpha^{3}N_{s}+3\alpha^{2}\beta N_{s}+3\alpha\beta^{2}N_{s}+\beta^{3}N_{s},
\end{equation}
which, for convenience, can be written as
\begin{eqnarray}
	(\alpha+\beta)^{4-1}N_{s}&=&\underbrace{\alpha^{4-1}\beta^{1-1}N_{s}}_{N_{(4,1)}}+\underbrace{3\alpha^{4-2}\beta^{2-1}N_{s}}_{N_{(4,2)}}\nonumber\\
	&+&\underbrace{3\alpha^{4-3}\beta^{3-1}N_{s}}_{N_{(4,3)}}+\underbrace{\alpha^{4-4}\beta^{4-1}N_{s}}_{N_{(4,4)}}.
	\label{con-3}
\end{eqnarray}
From equation (\ref{con-2}), we identify $N_{(4,1)}$, $N_{(4,2)}$, $N_{(4,3)}$, and $N_{(4,4)}$.

The same pattern found in equations (\ref{con-1}), (\ref{con-2}), and (\ref{con-3}) can be obtained
for $(\alpha+\beta)^{s}$, with $s=4,5,6...$.
Thus, from these equations, one can see that $N_{(i,j)}$ can be written in terms of binomial coefficients. It is well known that
\begin{equation}
	(\alpha+\beta)^{n}=\sum_{k=0}^{n}\frac{n!}{k!\left(n-k\right)!}\alpha^{k}\beta^{n-k}.
	\label{alpha-beta-n}
\end{equation}
Taking as basis equations (\ref{con-1}), (\ref{con-2}), and (\ref{con-3}), and making
\begin{eqnarray}
	n&=&i-1,\\
	k&=&i-j,
\end{eqnarray}
equation (\ref{alpha-beta-n}) can be written as
\begin{equation}
	(\alpha+\beta)^{i-1}=\sum_{j=1}^{i}\frac{\left(i-1\right)!}{\left(i-j\right)!
		\left(j-1\right)!}\alpha^{i-j}\beta^{j-1}.
	\label{ss1}
\end{equation}
Multiplying both sides of equation (\ref{ss1}) by $N_{s}$, and comparing
with equations (\ref{con-1}), (\ref{con-2}), and (\ref{con-3}), we can write
\begin{equation}
	(\alpha+\beta)^{i-1}N_{s}=\sum_{j=1}^{i}\underbrace{\frac{\left(i-1\right)!}{\left(i-j\right)!\left(j-1\right)!}\alpha^{i-j}\beta^{j-1}N_{s}}_{N_{(i,j)}},
\end{equation}
from which we identify $N_{(i,j)}$, and thus obtaining equation (\ref{N-teorico}).


\end{document}